\documentclass[twocolumn,english,aps,pra,showpacs,showkeys]{revtex4-1}
\usepackage[T1]{fontenc}
\usepackage[latin9]{inputenc}
\usepackage{amsmath}
\usepackage{graphicx}
\usepackage{amssymb}
\usepackage{amsfonts}
\usepackage{amsthm}
\usepackage{multirow}

\makeatletter

\@ifundefined{textcolor}{}
{%
 \definecolor{BLACK}{gray}{0}
 \definecolor{WHITE}{gray}{1}
 \definecolor{RED}{rgb}{1,0,0}
 \definecolor{GREEN}{rgb}{0,1,0}
 \definecolor{BLUE}{rgb}{0,0,1}
 \definecolor{CYAN}{cmyk}{1,0,0,0}
 \definecolor{MAGENTA}{cmyk}{0,1,0,0}
 \definecolor{YELLOW}{cmyk}{0,0,1,0}
 }

\usepackage{txfonts}

\makeatother

\usepackage{babel}

\begin{document}

\title{Quantum correlations in qutrit-qutrit systems under local quantum noise channels}

\author{Nasibollah Doustimotlagh}
\affiliation{State Key Laboratory of Low-Dimensional Quantum Physics and Department of Physics, Tsinghua University, Beijing 100084,
China}
\author{Jin-Liang Guo}
\affiliation{State Key Laboratory of Low-Dimensional Quantum Physics and Department of Physics, Tsinghua University, Beijing 100084,
China}
\author{Shuhao Wang}
\email[E-mail: ]{wsh@physics.so}
\affiliation{State Key Laboratory of Low-Dimensional Quantum Physics and Department of Physics, Tsinghua University, Beijing 100084,
China}

\date{\today }
\begin{abstract}
Due to decoherence, realistic quantum systems inevitably interact with the environment when quantum
information is processed, which causes the loss of quantum properties. As a fundamental issue of quantum properties, quantum correlations have attracted a lot of interests in recent years. Because of the importance of high dimensional systems in quantum information, in this work, we
study the quantum correlations affected by the Markovian environment by considering the quantum correlations of qutrit-qutrit quantum systems measured by the negativity and the geometric discord. 
The local noise channels covered in this work includes dephasing, trit-flip, trit-phase-flip, and depolarising channels.
We have also investigated the cases where the local decoherence channels of two sides are
identical and non-identical.
\end{abstract}
\pacs{03.67.YZ, 03.65.Ud}
\keywords{Quantum correlations, Decoherence, Quantum noise, Quantum channel}

\maketitle

\section{Introduction}
\label{intro}

Quantum correlations, which lie in the foundations
of quantum theory, have been of renewed interest during the last two decades as the
field of quantum information science emerged and matured. 
Quantum entanglement, as a kind of quantum correlations, has been playing central
roles in quantum information and computation \cite{bostr,yanf,dengf,yangy,shig}. The negativity is one of the best known and most popular tools of quantifying bipartite quantum entanglement. It can be computed easily for arbitrary states of a composite system. Therefore, it has received various studies \cite{Nakano,Eltschka,Datta1,Ferrie,Dajka1}. 
However, quantum entanglement does not account for
all of the non-classical properties of quantum phenomenons. Therefore, numerous quantifiers of quantum correlations have
been further proposed to reveal the non-classical correlations that cannot be fully captured by quantum entanglement \cite{Nielson,Rudolph,Moradi,Yu,Vidal1}. Among the various measures, it has been shown that quantum discord provides a larger region of quantum states with non-classical correlations, and a non-zero discord quantum state without entanglement may be responsible for the efficiency of a quantum computer. Therefore,
much attention has been paid to study the quantum discord in various quantum systems
\cite{Horodecki,Ollivier,Brodutch,Datta,Lanyon,Okrasaa,Zhang,Hej,Liu,Chen,Ali}. However, due to the complicated optimization, it is usually
complicated to calculate quantum discord analytically. Even for qubit-qubit quantum systems, the analytical formula of quantum discord can only be obtained for a few cases, and a
general method is still lacking. As a variant of quantum discord, geometric discord (GD) was introduced by Dakic and coworkers \cite{Dakic}. Due to its simplicity in calculation, it has attracted a lot of research interests \cite{Rana,Vidal}.

It is well known that quantum systems inevitably interact with the environment when quantum information is processed in the real-world, which brings about the loss of initially presented quantum properties  \cite{doosti,Wang,Jing,Hu,Miao,Ramzan}. This phenomenon can be characterized by decoherence, which can be illustrated by the example of faithfully transmitting an unknown quantum state through a noisy quantum channel. During the transmission, the carrier of the information interacts with the channel and gets correlated with other degrees of freedom. This gives rise to the phenomenon of so-called 'decoherence' on the subspace of the information carrier.

Recent studies on open quantum systems highlight the existence of two different classes of dynamical behaviours known as Markovian and non-Markovian regimes \cite{Ren,Guo}. In the quantum domain, under certain assumptions, Markovian dynamics lead to a master equation in the Lindblad form. However, in certain scenarios such approximations are not justified and one needs to go beyond perturbation theory. It is clear that, due to the general complexity of the problem to be studied, exact solutions exist only for simple open quantum systems models such as the well-known Jaynes-Cummings model, quantum Brownian motion model, and certain pure dephasing models \cite{Chou,An,Feng}. For the deeper understanding of quantum phenomenons, it is desirable to investigate the behaviours of the quantum properties under the action of decoherence. The previous work showed that quantum discord and entanglement behave differently under the effect of the environment \cite{Bylicka,Zou,Maziero}. In particular, for certain quantum states, the phenomenon of entanglement sudden death does not occur for quantum discord. There exists some studies on the quantum correlations of qubit-qubit and qubit-qutrit systems under the effect of local quantum noises in the Markovian environment \cite{Ren,Guo,Bylicka,Zou,Maziero,Fanchini,Girolami}. The importance of quantum states with higher dimensions
is gradually recognized in recent years \cite{Yao}. Therefore, a deep understanding of the quantum correlations dynamics of the high-dimensional quantum systems beyond qubits is very practical.

Song et al. \cite{Song} have studied the distillability in qutrit-qutrit systems under local dephasing channels, and found that the distillability will be decrease after a finite time.
M. Ali studied the distillability sudden death in qutrit-qutrit systems under global and multilocal dephasing \cite{Ali1}, and amplitude damping channels \cite{Ali2}. He has found that the quantum entanglement will decrease after a finite time under dicoherence. Guo et al. \cite{Guo} studied the dynamics of quantum correlations of qubit-qutrit systems under various decoherent channels. They have shown that the decoherent channels bring with different influences for the dynamics of quantum correlations measured by negativity, quantum discord and GD. The influences depend on the initial state parameters and the properties of the decoherent channels. In this work, we will study the quantum correlations in qutrit-qutrit systems under identical and non-identical local noise channels. We consider four kinds of local noise channels on each qutrit, including dephasing, trit-flip, trit-phase-flip, and depolarising channels. The remainder
of this paper is organized as follows:
In Sec. \ref{sec:2}, we will introduce the negativity and the GD as our measures of quantum correlations. In Sec. \ref{sec:3}, the evolution of the quantum correlations for a qutrit-qutrit system under the identical and non-identical local noise channels are discussed. Finally, we conclude in Sec. \ref{sec:4}.

\section{Basic concepts}
\label{sec:2}
In this section, we will describe the GD and the negativity in a brief manner.

\subsection{Negativity}
\label{n}
The negativity is a useful entanglement measure for arbitrary
bipartite state quantum systems \cite{Nakano,Eltschka,Ferrie,Dajka1,Vidal}, especially for free entangled states. For a given density matrix $\rho_{AB}$, its negativity will be
\begin{equation}
N(\rho_{AB})=\frac{||\rho_{AB}^{T_A}||-1}{2},
\end{equation}
where $\rho_{AB}^{T_A}$ is the partial transpose of $\rho_{AB}$ with respect to subsystem $A$ and $||\Lambda||^2=\rm Tr|\Lambda|=\rm Tr\sqrt{\Lambda^\dagger\Lambda}$ represents the trace norm. If $N(\rho_{AB})\geq 0$, then the bipartite state is free entangled.

\subsection{GD}
\label{gd}
Geometric measure of quantum discord in a bipartite state can be described as the distance of the state from the closest zero-discord state \cite{Dakic,Hassan,Jianwei,Hassan1}
\begin{equation}
{\rm GD}_{\rho_{AB}}=\min_{\Xi\in \Gamma_0}||\rho_{AB}-\Xi||^2,
\end{equation}
where $||.||$ is the Hilbert-Schmidt
norm and $\Gamma_0$ is the set of all zero discord states. For a general bipartite state of $d_1\times d_2$ dimensions, the density matrix can be written as
\begin{equation}
\rho_{AB}=\frac{1}{d_1d_2}[\mathbb{I}_{d_1}\times \mathbb{I}_{d_2}+\sum_k y_k \xi_k\otimes \mathbb{I}_{d_2}+\sum_l z_l \mathbb{I}_{d_1}\otimes \xi_l +\sum_{k,l} \xi_k\otimes \xi_l],
\end{equation}
where $\xi_k$ ($k=1,...,d_1^2-1$) and $\xi_l$ ($l=1,...,d_2^2-1$) are the generators of ${\rm SU}(d_1)$ and ${\rm SU}(d_2)$ groups (special unitary groups of dimensions $d_1$ and $d_2$, respectively), satisfying the relation $\rm Tr{(\xi_k \xi_l)}=2\delta_{kl}$. 

The lower bound \cite{Rana,Hassan} of the GD has been derived as
\begin{equation}
{\rm GD}_{\rho_{AB}} \geq \frac{4}{d_1^2 d_2} (||Y||^2+\frac{2}{n}||V||^2-\sum_{n=1}^{d_1^2-1}\lambda_n),
\end{equation}
where $||Y||^2={\rm Tr}(YY^\dagger)$ with Bloch vector $Y=(y_1,y_2,y_3)^T$ with elements $y_i=\frac{d_1}{2}{\rm Tr}(\rho_{AB}\xi_i\otimes \mathbb{I}_{d_2})$, matrix $V=(v_{kl})_{3\times 3}$ is the correlation matrix with elements $v_{kl}=\frac{d_1 d_2}{4}{\rm Tr}(\rho_{AB}\xi_k\otimes\xi_l)$ and $\lambda_n$ are the eigenvalues of the matrix $YY^T+\frac{2VV^T}{d_2}$ arranged in non-increasing order. We will use the lower bound as the estimation of the GD in the following.

\section{Quantum correlations under local noise channels }
\label{sec:3}
Let us consider a qutrit-qutrit system with each particle interacts independently with the local noise
channels. In this case, the evolution of a quantum state can be
described by the Lindblad equation, which can be written in terms of the Kraus operators as
\begin{equation}
\rho_{AB}(t)=\sum_{j=1}^3\sum_{i=1}^3 \mathcal{F}_j^B \mathcal{E}_i^A \rho_{AB}(0) \mathcal{E}_i^{A^\dagger} \mathcal{F}_j^{B^\dagger},
\end{equation}
where $\mathcal{E}_i^A$ and $\mathcal{F}_j^B$ are the Kraus operators characterizing the local noise channels on
$A$ and $B$, respectively. They satisfy $ \sum_{i}\mathcal{E}_i^A \mathcal{E}_i^{A^{\dagger}}=\mathbb{I} $ and $ \sum_i \mathcal{F}_i^B \mathcal{F}_i^{B^{\dagger}}=\mathbb{I}$ \cite{An,Feng}.

To study the dynamics of quantum correlations for a qutrit-qutrit system, 
we choose the Bell state
\begin{equation}
\rho_{AB}(0)=\frac{1}{3}(|00\rangle+|11\rangle+|22\rangle)(\langle 00|+\langle 11|+\langle 22|)
\end{equation}
as the initial quantum state.
In the following, we will study the dynamics of the quantum correlations of $\rho_{AB}(t)$ under the effect of four different noise channels.

\subsection{Identical local noise channels}
\label{idt}
\subsubsection{Dephasing channels}
Physically, dephasing corresponds to any process of losing coherence without the
exchange of energy. Under the action of dephasing noises, the off-diagonal elements of
the density matrix decay exponentially against time. For two local dephasing
noises on qutrits $A$ and $B$, the Kraus operators are
given by
\begin{eqnarray}
&&\mathcal F_1^B=\mathbb{I}_A\otimes
 M_B, \mathcal F_2^B=\mathbb{I}_A\otimes N_B, \mathcal F_3^B=\mathbb{I}_A\otimes T_B
  ,\nonumber\\
&&\mathcal E_1^A=
 M_A \otimes\mathbb{I}_B,
  \mathcal E_2^A=
 N_A\otimes\mathbb{I}_B,
\mathcal E_3^A=
T_A\otimes\mathbb{I}_B,
\end{eqnarray}
where 
\begin{eqnarray}
&&M_i=\begin{pmatrix}
  1 & 0 & 0 \\
  0 & \sqrt{1-\gamma_i} & 0 \\
  0  & 0  &  \sqrt{1-\gamma_i}  \\
  \end{pmatrix},
 N_i=\begin{pmatrix}
  0 & 0 & 0 \\
  0 & \sqrt{\gamma_i} & 0 \\
  0  & 0  & 0  \\
  \end{pmatrix},
 T_i=\begin{pmatrix}
  0 & 0 & 0 \\
  0 & 0 & 0 \\
  0  & 0  & \sqrt{\gamma_i}  \\
  \end{pmatrix},
\end{eqnarray}
$i=A$ or $B$, $ \gamma_A=1-\exp(-tq_A) $ and $ \gamma_B=1-\exp(-tq_B) $ with $ q_A$ and $ q_B$ denoting the decay rates of qutrits $A$ and $B$, respectively.

\subsubsection{Trit-flip channels}
For trit-flip channels of qutrits $A$ and $B$, the Kraus operators are
given by
\begin{eqnarray}
&&\mathcal F_1^B=\mathbb{I}_A\otimes
 M_B, \hspace{1em}\mathcal F_2^B=\mathbb{I}_A\otimes N_B,
 \mathcal F_3^B=\mathbb{I}_A\otimes T_B,\nonumber\\
 &&\mathcal E_1^A=
 M_A \otimes\mathbb{I}_B,
    \mathcal E_2^A=
 N_A\otimes\mathbb{I}_B,
\mathcal E_3^A=
T_A\otimes\mathbb{I}_B,
  \end{eqnarray}
where
\begin{eqnarray}
&&M_i= \sqrt{1-2/3\gamma_i}\mathbb{I}_i,
 N_i=\sqrt{\gamma_i}\begin{pmatrix}
  0 & 0 & 1 \\
  1 & 0 & 0 \\
  0  & 1  & 0  \\
  \end{pmatrix},
  T_i=\sqrt{\gamma_i}\begin{pmatrix}
  0 & 1 & 0 \\
  0 & 0 & 1 \\
  1  & 0  & 0  \\
  \end{pmatrix}.
\end{eqnarray}

\subsubsection{Trit-phase-flip channels}
The Kraus operators are
\begin{eqnarray}
&&\mathcal F_1^B=\mathbb{I}_A\otimes
 M_B, \hspace{1em}\mathcal F_{2,3}^B=\mathbb{I}_A\otimes N_B^{\pm},
 \mathcal F_{4,5}^B=\mathbb{I}_A\otimes T_B^{\pm},\nonumber\\
 &&\mathcal E_1^A=
 M_A \otimes\mathbb{I}_B,
    \mathcal E_{2,3}^A=
 N_A^{\pm}\otimes\mathbb{I}_B,
\mathcal E_{4,5}^A=
T_A^{\pm}\otimes\mathbb{I}_B,
  \end{eqnarray}
where
\begin{eqnarray}
&&M_i= \sqrt{1-2/3\gamma_i}\mathbb{I}_i,\nonumber\\
 &&N_i^{\pm}=\sqrt{\gamma_i/6}\begin{pmatrix}
  0 & 0 & \exp (\pm i2\pi/3) \\
  1 & 0 & 0 \\
  0  & \exp (\mp i2\pi/3)  & 0  \\
  \end{pmatrix},\nonumber\\
 && T_i^{\pm}=\sqrt{\gamma_i/6}\begin{pmatrix}
  0 & \exp (\mp i2\pi/3) & 0 \\
  0 & 0 & \exp (\pm i2\pi/3) \\
  1  & 0  & 0  \\
  \end{pmatrix}.
\end{eqnarray}

\begin{figure*}[]
\begin{centering}
\includegraphics[width=12.4cm]{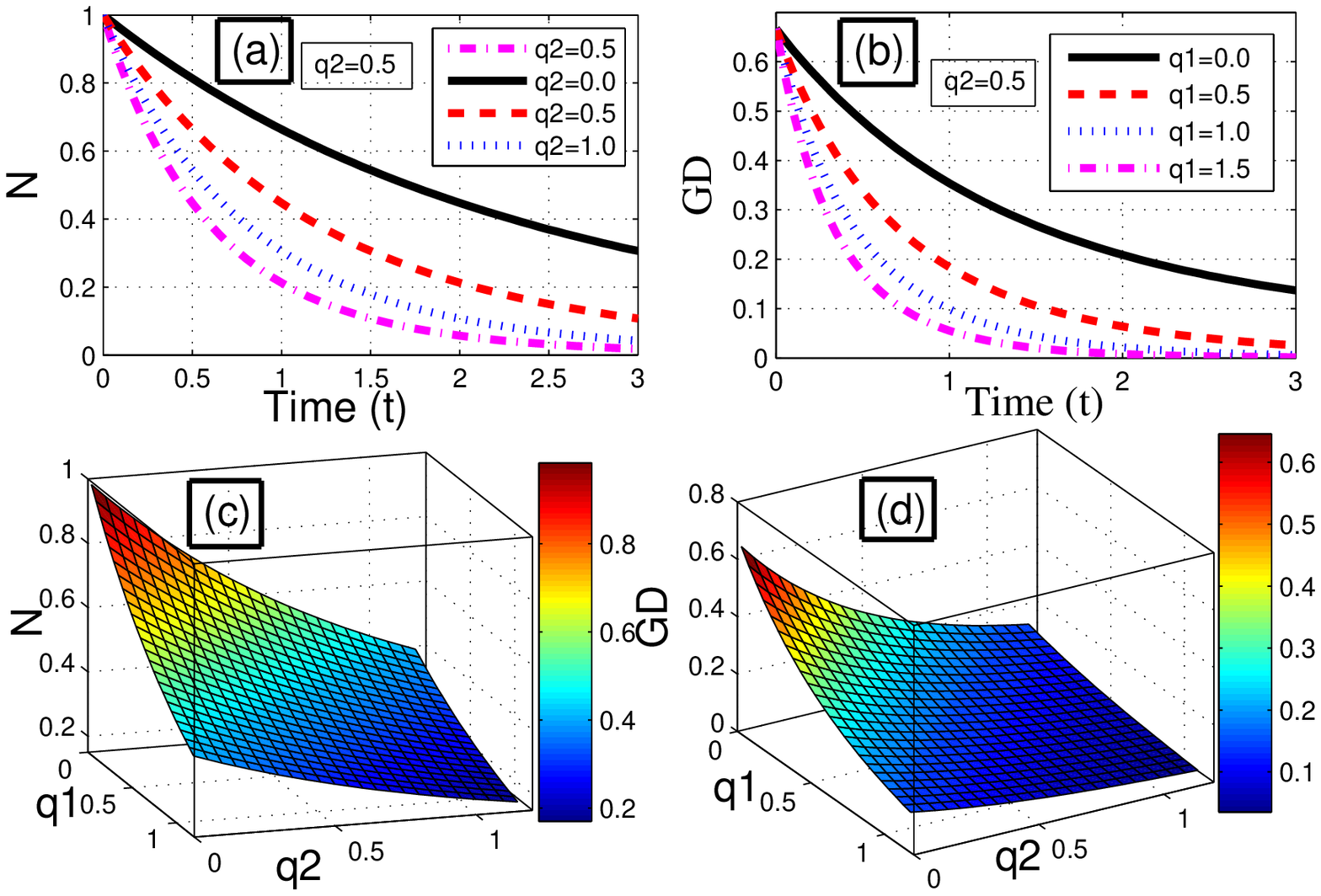}
\caption{(Color online)  The negativity (signed by N) and the GD of a qutrit-qutrit quantum systems under the effect of two identical phase-flip channels.  (a) The negativity and (b) the GD, where the decay rate for one channel is fixed ($q_2=0.5$).(c) The negativity and (d) the GD when the time is fixed ($t=1.0$).}
\label{fig1}
\end{centering}
\end{figure*}

\begin{figure*}[]
\begin{centering}
\includegraphics[width=12.4cm]{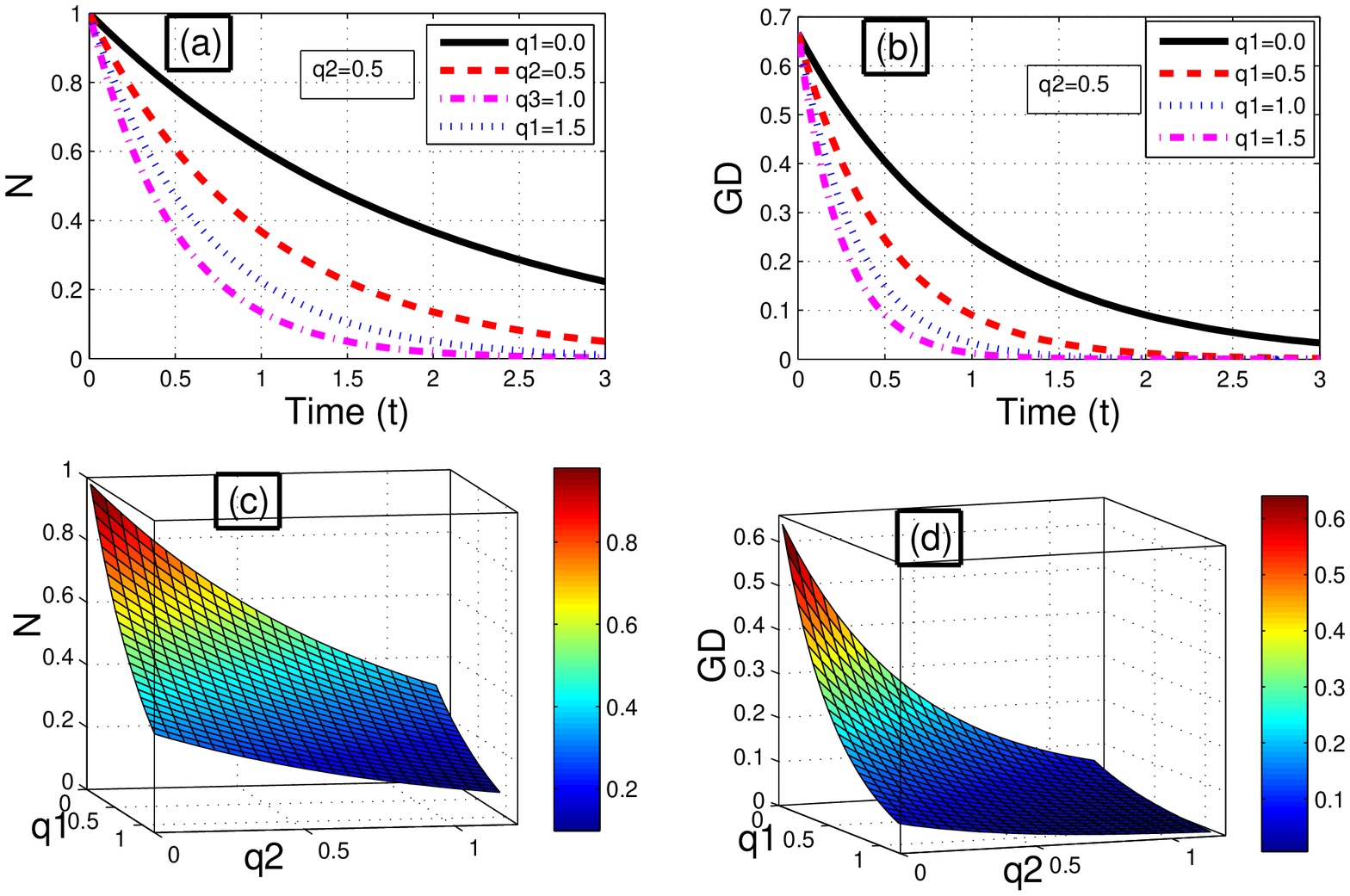}
\caption{(Color online) The negativity and the GD of a qutrit-qutrit quantum systems under the effect of two identical trit flip channels.  (a) The negativity and (b) the GD, where the decay rate for one channel is fixed ($q_2=0.5$).(c) The negativity and (d) the GD when the time is fixed ($t=1.0$).}
\label{fig2}
\end{centering}
\end{figure*}
 \begin{figure*}[]
\begin{centering}
\includegraphics[width=12.4cm]{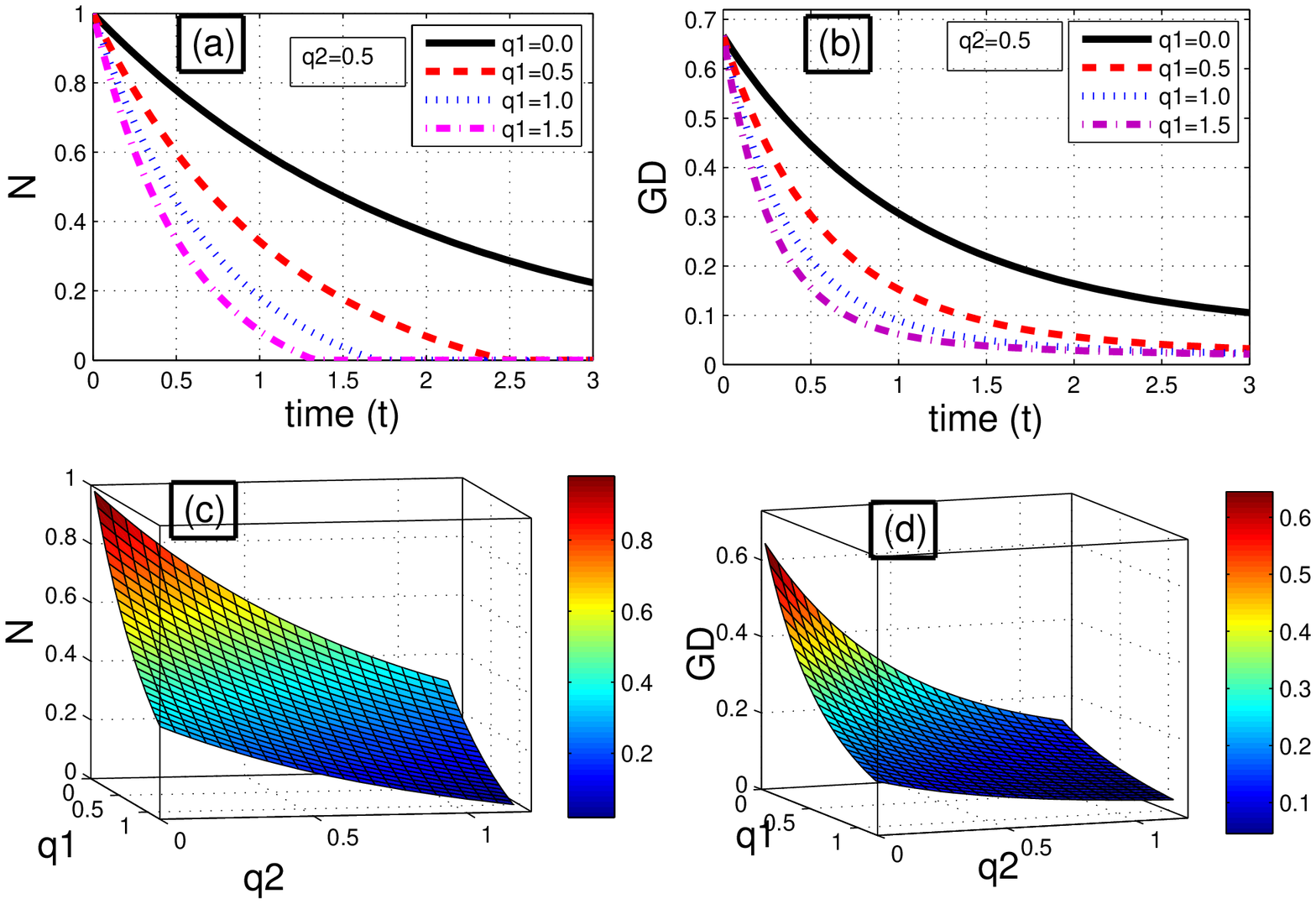}
\caption{(Color online)  The negativity and the GD of a qutrit-qutrit quantum systems under the effect of two identical trit-phase-flip channels.  (a) The negativity and (b) the GD, where the decay rate for one channel is fixed ($q_2=0.5$).(c) The negativity and (d) the GD when the time is fixed ($t=1.0$).}
\label{fig3}
\end{centering}
\end{figure*}
 \begin{figure*}[]
\begin{centering}
\includegraphics[width=12.4cm]{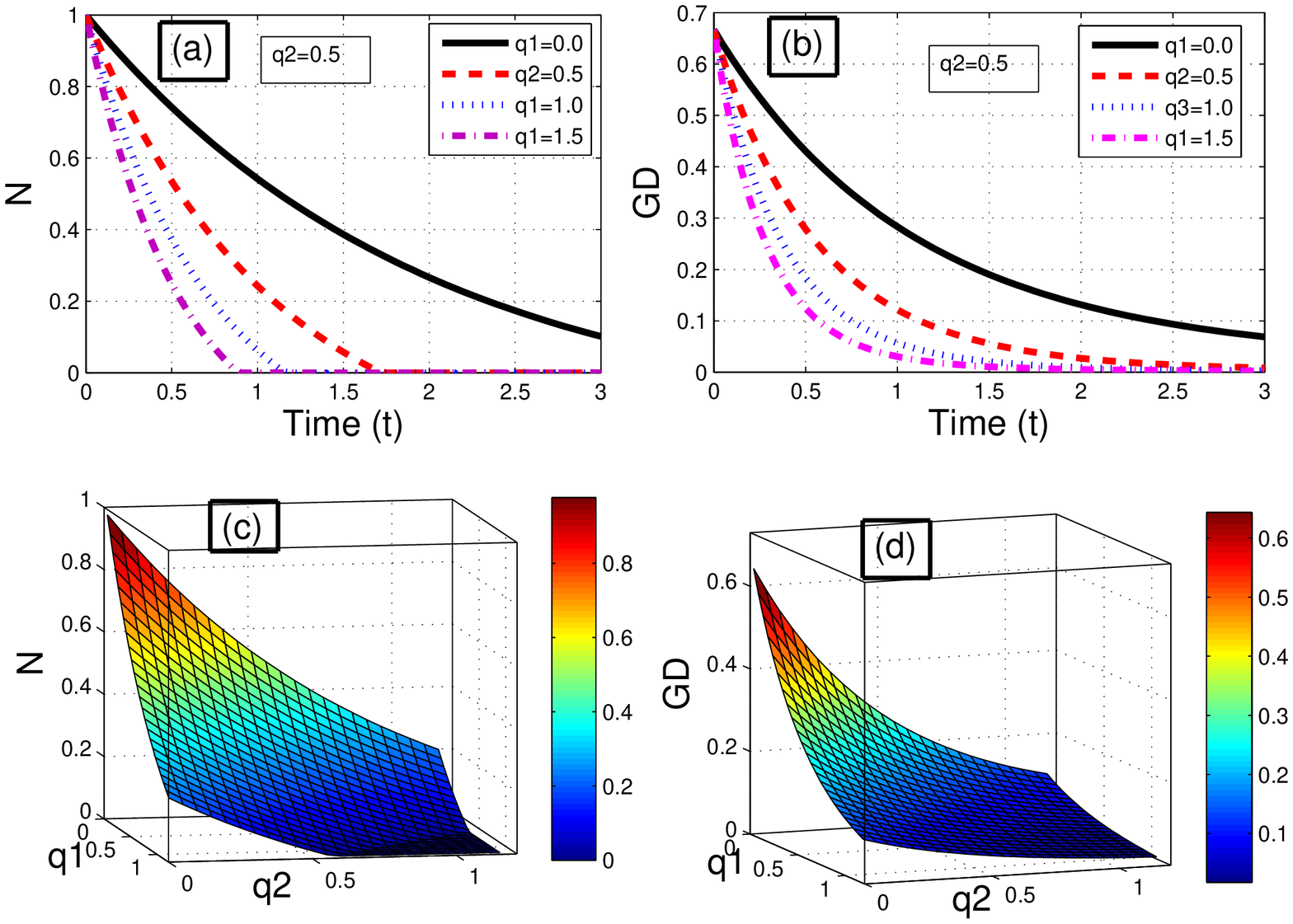}
\caption{(Color online)  The negativity and the GD of a qutrit-qutrit quantum systems under the effect of two identical depolarizing channels.  (a) The negativity and (b) the GD, where the decay rate for one channel is fixed ($q_2=0.5$).(c) The negativity and (d) the GD when the time is fixed ($t=1.0$).}
\label{fig4}
\end{centering}
\end{figure*}
\subsubsection{Depolarising channels}
The Kraus operators are
\begin{eqnarray}
&&\mathcal F_1^B=\mathbb{I}_A\otimes \alpha_B\mathbb{I}_B,
\mathcal F_2^B=\mathbb{I}_A\otimes \beta_B\mathbb{I}_B,
\mathcal F_3^B=\mathbb{I}_A\otimes \beta_BZ,\nonumber\\
 &&\mathcal F_4^B=\mathbb{I}_A\otimes \beta_BY^2,
\mathcal F_5^B=\mathbb{I}_A\otimes \beta_BYZ,
\mathcal F_6^B=\mathbb{I}_A\otimes \beta_BY^2Z,\nonumber\\
 &&\mathcal F_7^B=\mathbb{I}_A\otimes \beta_BYZ^2,
\mathcal F_8^B=\mathbb{I}_A\otimes \beta_BY^2Z^2,
\mathcal F_9^B=\mathbb{I}_A\otimes \beta_BZ^2,\nonumber\\
 &&\mathcal E_1^A= \alpha_A\mathbb{I}_A\otimes\mathbb{I}_B,
 \mathcal E_2^A= \beta_A\mathbb{I}_A\otimes\mathbb{I}_B,
 \mathcal E_3^A= \beta_AZ\otimes\mathbb{I}_B, \nonumber\\
 &&\mathcal E_4^A= \beta_AY^2\otimes\mathbb{I}_B,
 \mathcal E_5^A= \beta_AYZ\otimes\mathbb{I}_B,
 \mathcal E_6^A= \beta_AY^2Z\otimes\mathbb{I}_B, \nonumber\\
 &&\mathcal E_7^A= \beta_AYZ^2\otimes\mathbb{I}_B,
 \mathcal E_8^A= \beta_AY^2Z^2\otimes\mathbb{I}_B,
 \mathcal E_9^A= \beta_AZ^2\otimes\mathbb{I}_B,
  \end{eqnarray}
  where
  \begin{eqnarray}
  &&
 \alpha_i= \sqrt{1-8/9\gamma_i},\hspace{2em}\beta_i= \sqrt{\gamma_i}/3,\nonumber\\
 &&Y=\begin{pmatrix}
  0 & 1 & 0 \\
  0 & 0 & 1 \\
  1  & 0  & 0  \\
  \end{pmatrix},
  Z=\begin{pmatrix}
  0 & 1 & 0 \\
  0  & \exp(i2\pi/3)& 0 \\
  1  & 0  & \exp(-i2\pi/3)  \\
  \end{pmatrix}.
  \end{eqnarray}

\subsubsection{Results}
 
Based on the numerical simulations, we give our results in Figs. \ref{fig1}, \ref{fig2}, \ref{fig3}, and \ref{fig4} for each two identical dephasing, trit-flip, trit-phase-filp, and depolarizing channels, respectively. One can observe that under the effect of two identical noise channels, as $q_1$ or $q_2$ increases, the negativity and the GD decrease rapidly (see $(b)$ and $(c)$ in the figures), which indicates that quantum correlation is strongly affected by large decay rates \cite{Ren,Guo}. By observing Figs. \ref{fig1} and \ref{fig2}, one can see that the GD is more robust than the negativity ageist two identical dephasing and trit-flip noise channels. However, as shown in Figs. \ref{fig3} and \ref{fig4}, for two identical tirt-phase-flip and depolarizing channels, the negativity becomes more robust than the GD. To study the effect of the decay rates of the channels, we draw the negativity and the GD versus the decay rates $q_1$ or $q_2$ in three-dimension figures with a fixed time $(t=1)$ (see $(b)$ and $(c)$ in the figures). It can been seen that in the case where $q_1$ and $q_2$ are large, the negativity and the GD are influenced strongly during the decoherence.

\subsection{Non-identical local noise channels}
In this section, we discuss the negativity and the GD under non-identical local noise channels, namely, the
noise channels are of different types and decay rates. The kraus operators have been given in Sec. \ref{idt}.

\begin{figure*}[]
\begin{centering}
\includegraphics[width=12.4cm]{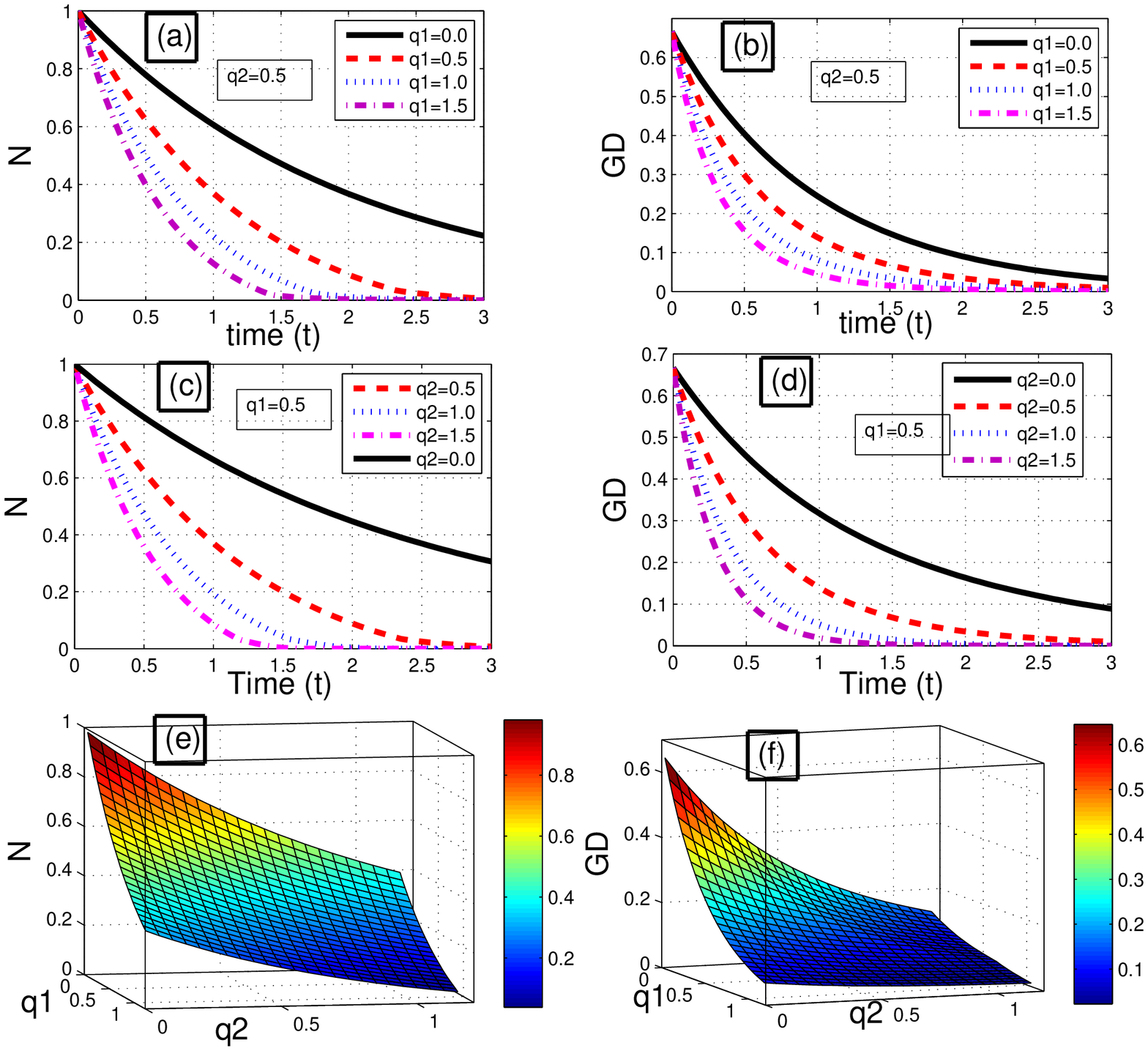}
\caption{(Color online) The negativity and the GD of a qutrit-qutrit quantum systems under the effect of two different noise channels such that first channel is dephasing and second one is trit-flip. (a) The negativity and (b) the GD, where the decay rate for second channel is fixed ($q_2=0.5$).(c) The negativity and (d) the GD, where the decay rate for first channel is fixed ($q_2=0.5$). (e) The negativity and (f) the GD when the time is fixed ($t=1.0$).}
\label{fig5}
\end{centering}
\end{figure*}

\begin{figure*}[]
\begin{centering}
\includegraphics[width=12.4cm]{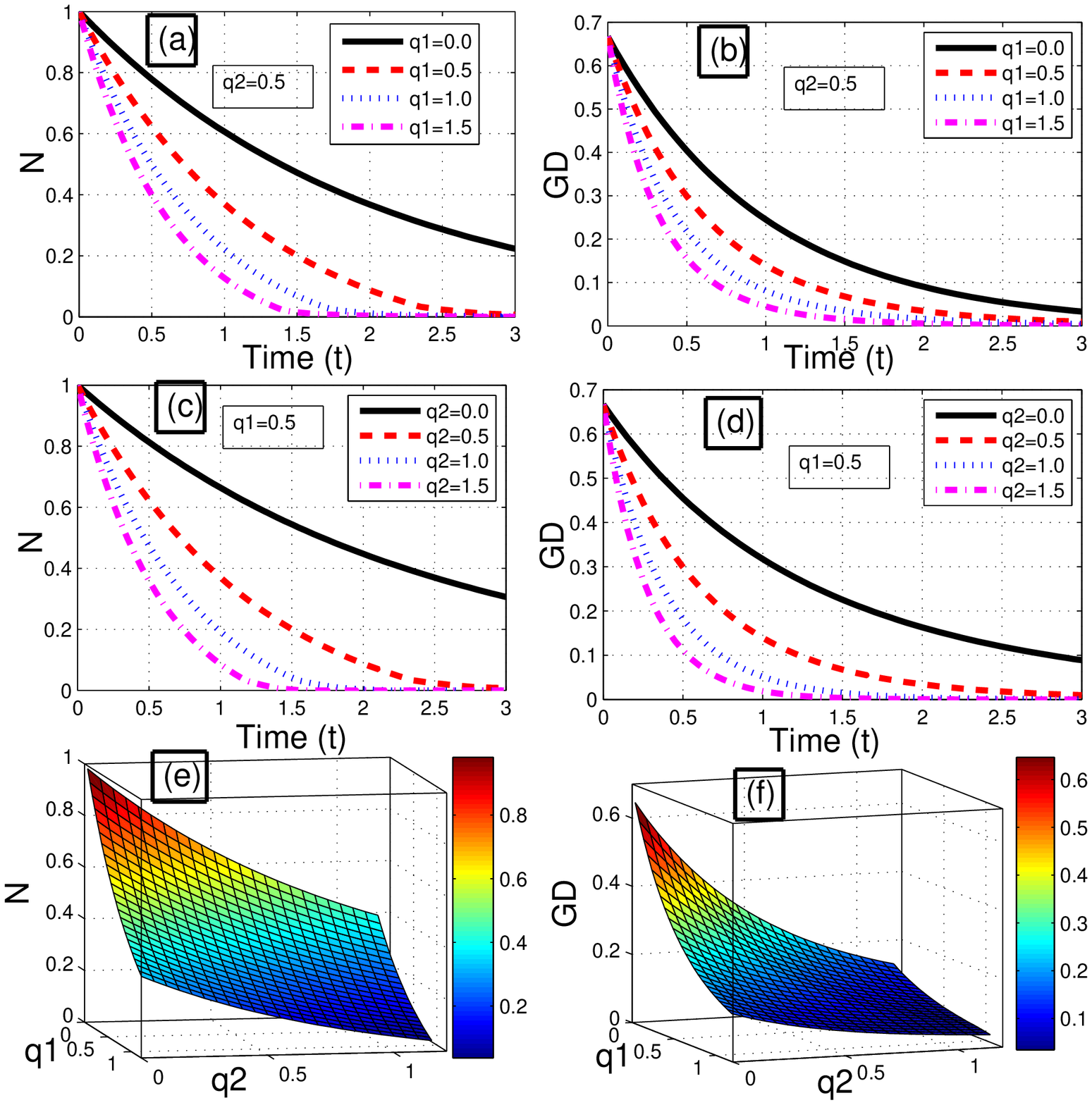}
\caption{(Color online) The negativity and the GD of a qutrit-qutrit quantum systems under the effect of two different noise channels such that first channel is dephasing and second one is trit-phase-flip. (a) The negativity and (b) the GD, where the decay rate for second channel is fixed ($q_2=0.5$).(c) The negativity and (d) the GD, where the decay rate for first channel is fixed ($q_2=0.5$). (e) The negativity and (f) the GD when the time is fixed ($t=1.0$).}
\label{fig6}
\end{centering}
\end{figure*}

\begin{figure*}[]
\begin{centering}
\includegraphics[width=12.4cm]{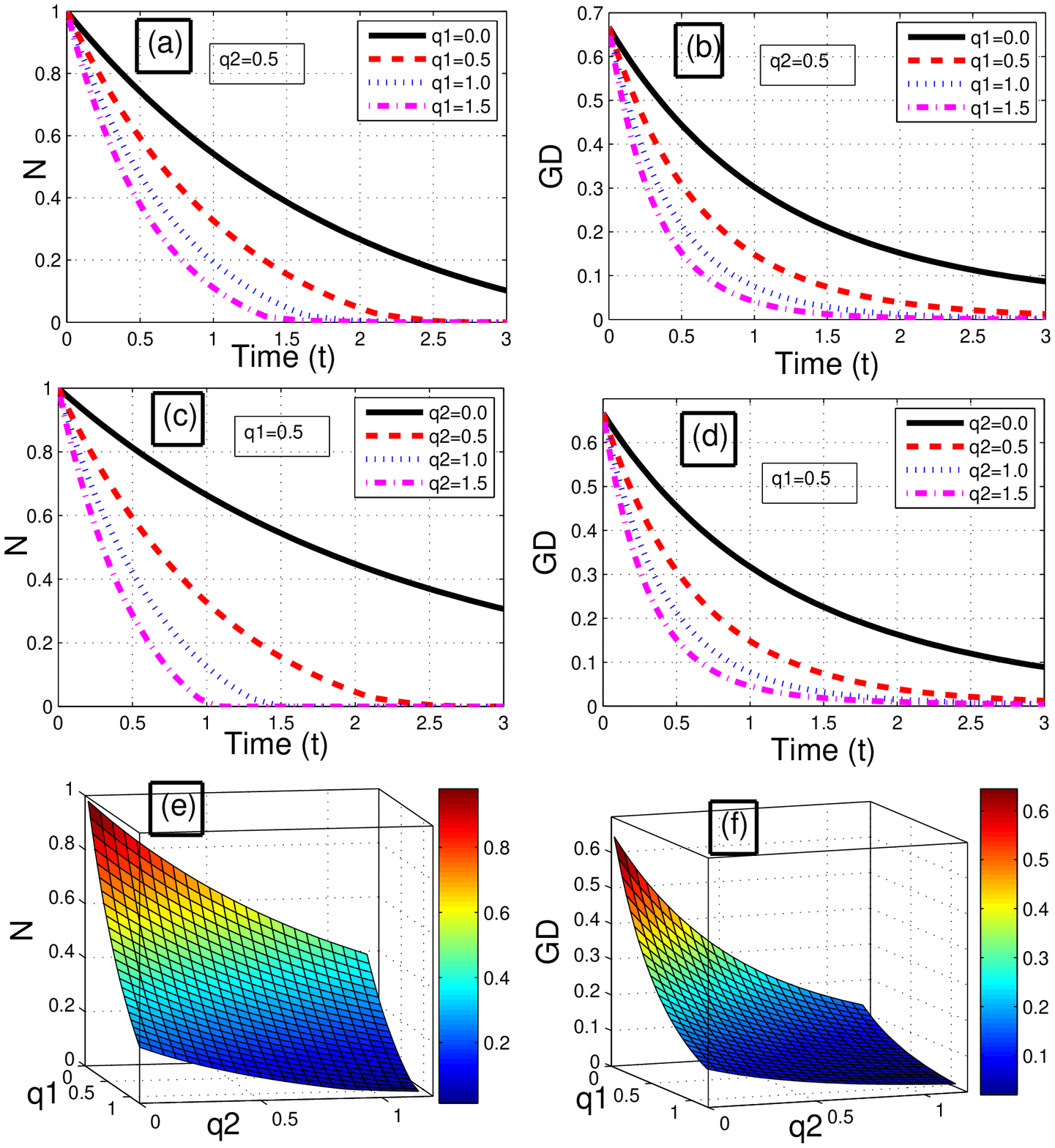}
\caption{(Color online) The negativity and the GD of a qutrit-qutrit quantum systems under the effect of two different noise channels such that first channel is dephasing and second one is depolarizing . (a) The negativity and (b) the GD, where the decay rate for second channel is fixed ($q_2=0.5$).(c) The negativity and (d) the GD, where the decay rate for first channel is fixed ($q_2=0.5$). (e) The negativity and (f) the GD when the time is fixed ($t=1.0$).}
\label{fig7}
\end{centering}
\end{figure*}

\begin{figure*}[]
\begin{centering}
\includegraphics[width=12.4cm]{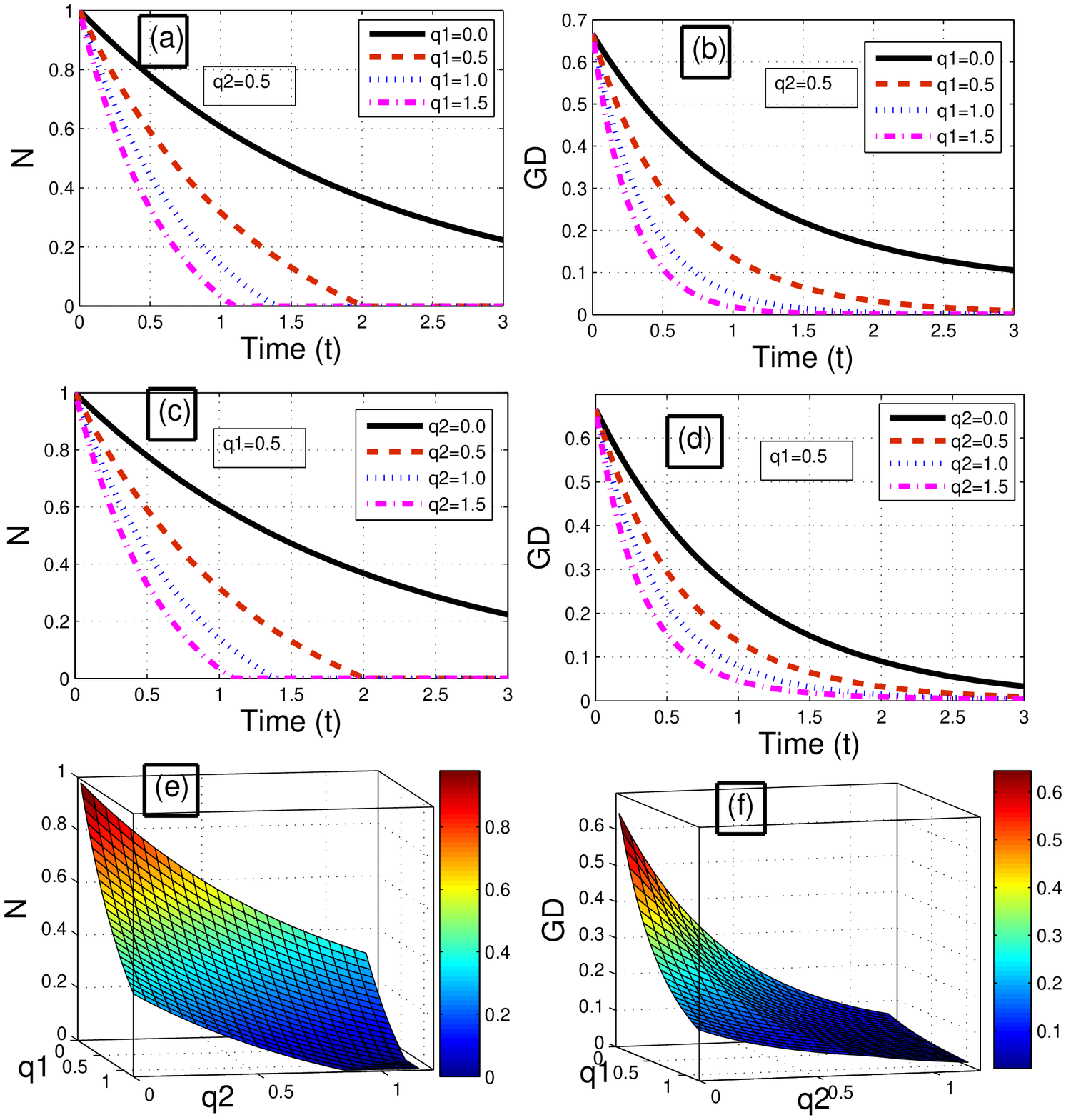}
\caption{(Color online) The negativity and the GD of a qutrit-qutrit quantum systems under the effect of two different noise channels such that first channel is trit-flip and second one is trit-phase-flip. (a) The negativity and (b) the GD, where the decay rate for second channel is fixed ($q_2=0.5$).(c) The negativity and (d) the GD, where the decay rate for first channel is fixed ($q_2=0.5$). (e) The negativity and (f) the GD when the time is fixed ($t=1.0$).}
\label{fig8}
\end{centering}
\end{figure*}

\begin{figure*}[]
\begin{centering}
\includegraphics[width=12.4cm]{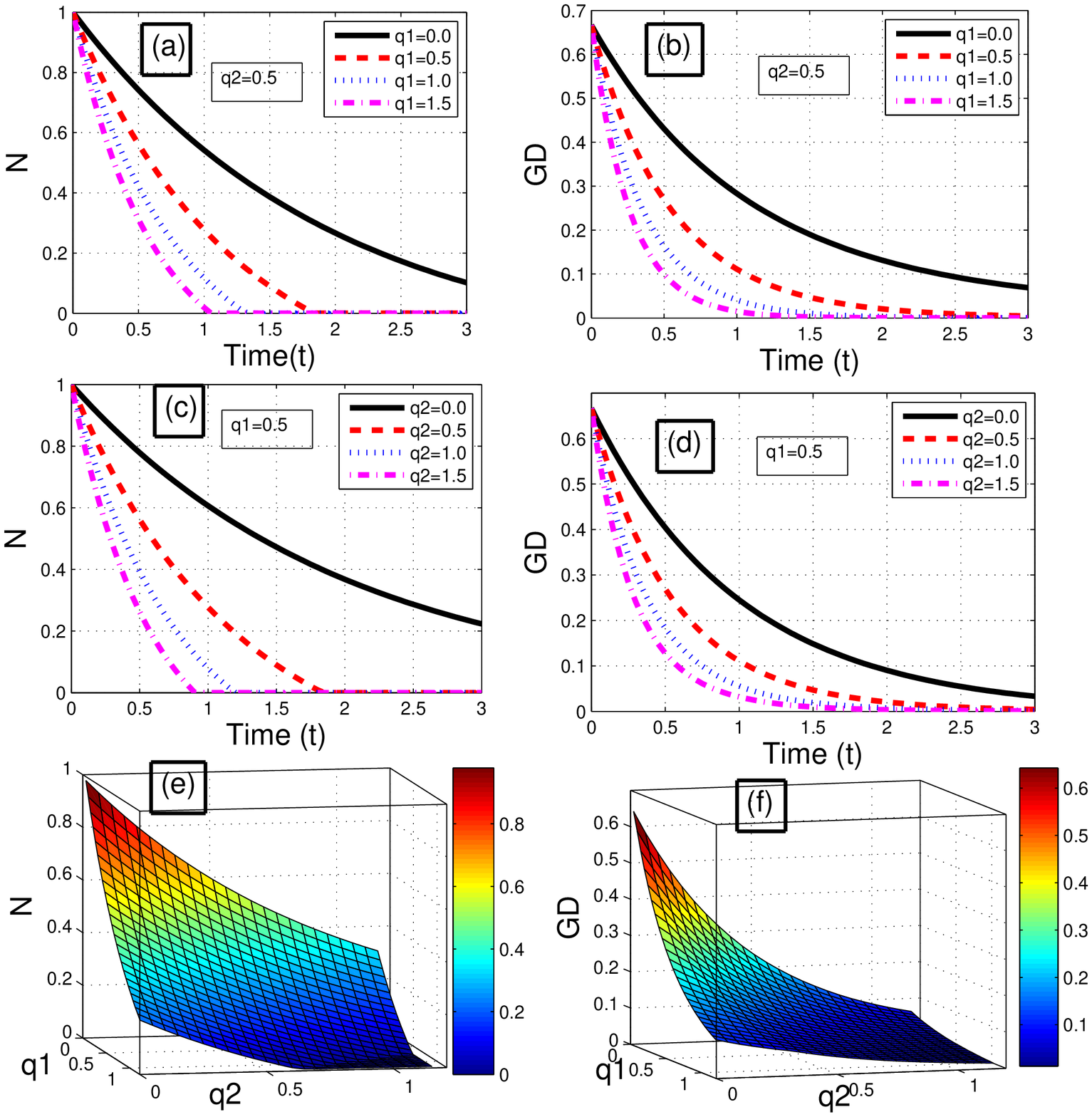}
\caption{(Color online) The negativity and the GD of a qutrit-qutrit quantum systems under the effect of two different noise channels such that first channel is trit-flip and second one is depolarizing. (a) The negativity and (b) the GD, where the decay rate for second channel is fixed ($q_2=0.5$).(c) The negativity and (d) the GD, where the decay rate for first channel is fixed ($q_2=0.5$). (e) The negativity and (f) the GD when the time is fixed ($t=1.0$).}
\label{fig9}
\end{centering}
\end{figure*}

\begin{figure*}[]
\begin{centering}
\includegraphics[width=12.4cm]{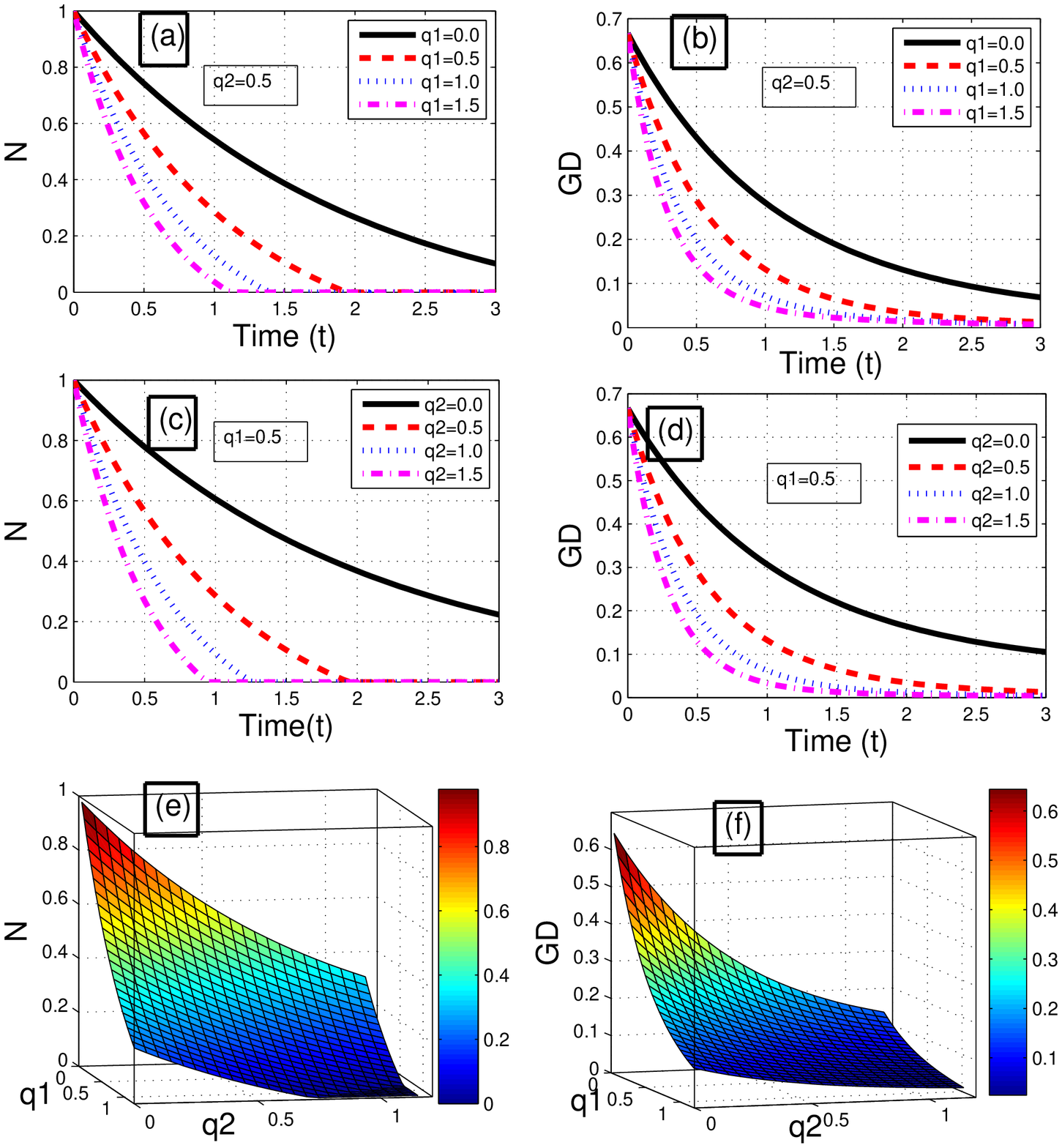}
\caption{(Color online) The negativity and the GD of a qutrit-qutrit quantum systems under the effect of two different noise channels such that first channel is trit-phase-flip and second one is depolarizing. (a) The negativity and (b) the GD, where the decay rate for second channel is fixed ($q_2=0.5$).(c) The negativity and (d) the GD, where the decay rate for first channel is fixed ($q_2=0.5$). (e) The negativity and (f) the GD when the time is fixed ($t=1.0$).}
\label{fig10}
\end{centering}
\end{figure*}

The results are drawn in Figs. \ref{fig5}, \ref{fig6}, \ref{fig7}, \ref{fig8}, \ref{fig9}, and \ref{fig10} for dephasing - trit flip channels, dephasing - trit phase flip channels, dephasing - depolarising channels, trit flip - trit phase flip channels, trit flip - depolarising channels, and  trit phase flip - depolarising channels respectively. It can be seen that under the effect of two different local noise channels, the negativity and the GD behave similar to the case where the channels are identical.
One can observe that in all cases of non-identical local noise channels the GD is more robust than the negativity. Same in the identical channels, as the time elapsed the state losses its quantum entanglement. To study just the effect of rate of the channels we draw the negativity and the GD versus the rates of two channels in three dimensions at a fixed time $(t=1)$ (see $(e)$ and $(f)$ in the figures). As the rates of the channels are large, the negativity and the GD decease rapidly.


\section{Conclusion}
\label{sec:4}

To summarize, in this work, we have studied the quantum correlations of a qutrit-qutrit quantum system under identical and non-identical local noise channels by using the negativity and the GD. We have considered four kinds of local noise channels on each qutrit, including dephasing, trit-flip, trit-phase-flip, and depolarising channels.
Meanwhile, we have further
investigated the effect of various combinations local noise channels on the quantum correlation. The results show that the quantum correlations decrease monotonically with time under decoherence. Meanwhile, when the time is fixed, negativity and geometric discord decrease rapidly when the decay rates are large. We expect our work may find further theoretical and experimental applications.


\begin{thebibliography}{99}

\bibitem{bostr} Bostrom, K., Felbinger, T.: Phys. Rev. Lett. {\bf 89}, 187902 (2002).
\bibitem{yanf} YU, G.: Int. J. Theor. Phys. {\bf 51}, 2954 (2012).
\bibitem{dengf} Deng, F. G., Li, X. H., at all.: Phys. Lett. A {\bf 72}, 359 (2006).
\bibitem{yangy} Yang, Y. G., Wen, Q. Y.: Sci. China Ser. G 5{\bf 50}, 558 (2007).
\bibitem{shig} Guhne, A., Hyllus, P.: Int. J. Theor. Phys. {\bf 42}, 1001 (2003).
\bibitem{Nakano} Nakano, T., Piani, M., Adesso, G.: Phys. Rev. A {\bf 72}, 012117 (2013)
\bibitem{Eltschka} Eltschka, C., Siewert, J.: Phys. Rev. Lett. {\bf 111}, 100503 (2013).
\bibitem{Datta1} Datta, A.: Phys. Rev. A {\bf 81}, 052312 (2010).
\bibitem{Ferrie} Ferrie, C., Morris, R., Emerson, J.: Phys. Rev. A {\bf 82}, 044103 (2010).
\bibitem{Dajka1} Dajka, J., Mierzejewski, M., £uczka, J., Blattmann, R., Hanggi, P.: J. Phys. A: Math. Theor. {\bf 45}, 485306 (2012).
\bibitem{Nielson} Nielson, M.A., Chuang, I.L.: Cambridge University Press, Cambridge (2000).
\bibitem{Rudolph} Rudolph, O.: Phys. Rev. A {\bf 67}, 032312 (2003).
\bibitem{Moradi} Moradi, S.: Quant. Inf. Comp. {\bf 11}, 957 (2011).
\bibitem{Yu} Yu, C.S., Song, H.S.: Phys. Rev. A {\bf 72}, 022333 (2005).
\bibitem{Vidal1} Vidal, G., Werner, R.F.: Phys. Rev. A {\bf 65}, 032314-032317 (2002).
\bibitem{Horodecki} Horodecki, R., Horodecki, P., Horodecki, M., Horodecki, K.: Rev. Mod. Phys. {\bf 81}, 865 (2009)
\bibitem{Ollivier} Ollivier, H,. Zurek, W.H.: Phys. Rev. Lett. {\bf 88}, 017901 (2001).
\bibitem{Brodutch}  Brodutch, A,. Terno, D.R.: Phys. Rev. A {\bf 83}, 010301(R) (2011).
\bibitem{Datta} Datta, A., Shaji, A., Caves, C.M.: Phys. Rev. Lett. {\bf 100}, 050502 (2008).
\bibitem{Lanyon}  Lanyon, B.P., Barbieri, M., Almeida, M.P., White, A.G.: Phys. Rev. Lett. {\bf 101}, 200501 (2008).
\bibitem{Okrasaa} Okrasaa, M.,Walczak, Z.:  Euro. phys. Lett. {\bf 96}, 60003 (2011).
\bibitem{Zhang} Yu. C.S., Zhang. Y., Zhao. H.: Quantum Inf. Process. (published online).
\bibitem{Hej} He, J., Tao Liu., W.: Int. J. Theor. Phys. {\bf 52}, 3381 (2013).
\bibitem{Liu} Liu, X., Ma, J., Xi, Z.,Wang, X.: Phys. Rev. A {\bf 83}, 012327 (2011).
\bibitem{Chen} Chen, Q., Zhang, C., Yu, S., Yi, X.X., Oh, C.H.: Phys. Rev. A {\bf 84}, 042313 (2011).
\bibitem{Ali} Ali, M.: J. Phys. A: Math. Theor. {\bf 43}, 495303 (2010).
\bibitem{Dakic} Dakic, B., Vedral, V., Brukner, C.: Phys. Rev. Lett. {\bf 105}, 190502 (2010).
\bibitem{Rana} Rana, S., Parashar, P.: Phys. Rev. A {\bf 85}, 024102 (2012).
\bibitem{Vidal} Vidal, G., Werner, R. F.: Phys. Rev. A {\bf 65}, 032314 (2002).
\bibitem{doosti} Doustimotlagh, N., Wang, S., You, C., Long, G. L.: EPL, {\bf 106}, 60003 (2014).
\bibitem{Wang} Wang, J., Jing, J.: Phys. Rev. A {\bf 82}, 032324 (2010).
\bibitem{Jing} Wang, J., Jing, J.: Ann. Phys. {\bf 327}, 283 (2012).
\bibitem{Hu} Hu, M.L., Fan, H.: Ann. Phys. {\bf 327}, 851 (2012).
\bibitem{Miao} Miao, C., Yang, M., Cao, Z, L.: Int. J. Theor. Phys. {\bf 52}, 1780 (2013).
\bibitem{Ramzan} Ramzan, M., Khan, M.K.: Quant. Inf. Process. {\bf 11}, 443 (2012).
\bibitem{Ren} Ren, B.C., Wei, H.R., Deng, F.G.: Quantum Inf. Process. {\bf 13}, 1175 (2014).
\bibitem{Guo} Guo, J.L., Li. H., Long. G. L.: Quantum Inf. Process. {\bf 12}, 3421 (2013).
\bibitem{Chou} Chou, C.H.,Yu, T., Hu,B.L.: Phys. Rev. E {\bf 77}, 011112 (2008).
\bibitem{An} An, J.H., Zhang,W.M.: Phys. Rev. A {\bf 76}, 042127 (2007).
\bibitem{Feng} An, J.H., Feng, M., Zhang, W.M.: Quantum Inf. Comput. {\bf 9}, 0317 (2009).
\bibitem{Bylicka} Addis. C., Bylicka.B., Chruscinski. D., Maniscalco. S.: arXiv:1402.4975v3.
\bibitem{Zou}Li, J.G., Zou, J., Shao, B.: Phys. Rev. A {\bf 82}, 042318 (2010).
\bibitem{Maziero} Maziero, J., Céleri, L.C., Serra, R.M., Vedral, V.: Phys. Rev. A {\bf 80}, 044102 (2009).
\bibitem{Fanchini} Fanchini, F.F.,Werlang, T., Brasil, C.A., Arruda, L.G.E., Caldeira, A.O.: Phys. Rev. A {\bf 81}, 052107 (2010).
\bibitem{Girolami} Girolami. D.: arXiv:1403.2446v1.
\bibitem{Yao} Wang, S.,Yao, L., Long, G.L.:Phys. Rev. A {\bf 87}, 062305 (2013).
\bibitem{Song} Song, W., Chen, L., Zhu, S, L.: Phys. Rev. A {\bf 80}, 012331 (2009).
\bibitem{Ali1} Mazhar, A.: Phys. Rev. A {\bf 81}, 042303 (2010).
\bibitem{Ali2} Mazhar, A.: J. Phys. B: At. Mol. Opt. Phys {\bf 43}, 045504 (2010).
\bibitem{Hassan} Hassan, A.S.M., Lari, B., Joag, P.S.: Phys. Rev. A {\bf 85}, 024302 (2012).
\bibitem{Jianwei} Xu, J.: J. Phys. A: Math. Theor. {\bf 45}, 405304 (2012).
\bibitem{Hassan1}Hassan, A.S.M., Joag, P.S.: J. Phys. A: Math. Theor. {\bf 45}, 345301 (2012).

\end{thebibliography}
\end{document}